# A Neural-Network-Embedded Equivalent Circuit Model for Lithium-ion Battery State Estimation

Zelin Guo, *Student Member, IEEE*, Yiyan Li*, *Member, IEEE*, Zheng Yan, *Member, IEEE*, Mo-Yuen Chow, *Fellow, IEEE*

*Abstract*—Equivalent Circuit Model (ECM) has been widely used in battery modeling and state estimation because of its simplicity, stability and interpretability. However, ECM may generate large estimation errors in extreme working conditions such as freezing environment temperature and complex charging/discharging behaviors, in which scenarios the electrochemical characteristics of the battery become extremely complex and nonlinear. In this paper, we propose a hybrid battery model by embedding neural networks as "virtual electronic components" into the classical ECM to enhance the model nonlinear-fitting ability and adaptability. First, the structure of the proposed hybrid model is introduced, where the embedded neural networks are targeted to fit the residuals of the classical ECM. Second, an iterative offline training strategy is designed to train the hybrid model by merging the battery state space equation into the neural network loss function. Last, the battery online state of charge (SOC) estimation is achieved based on the proposed hybrid model to demonstrate its application value. Simulation results based on a real-world battery dataset show that the proposed hybrid model can achieve 29% - 64% error reduction for SOC estimation under different operating conditions at varying environment temperatures.

## I. INTRODUCTION

STATE estimation of lithium-ion batteries, such as State of Charge (SOC) and State of Health (SOH) in electric vehicles [1] and State of Energy (SOE) in energy storage systems [2], plays a vital role in the Battery Management System (BMS) to maximize the system performance and economical values [3]. In general, the battery state estimation methods can be categorized into physics-based methods and data-driven methods [4].

Physics-based methods model the internal physical process of the lithium-ion batteries, such as the equivalent circuit model (ECM) and the electrochemical model [5]. Then the model parameters can be identified from field measurement data, based on which the battery state estimation can be further achieved. ECM has been widely studied and implemented in industry due to its simplicity, stability and interpretability, and a series of ECMs have been proposed such as the first-order model [6], second-order model [7], and fractional-order model [8]. [9] uses two Generalized Super-Twisting (GST) identification algorithms to identify the resistance and capacity of the first-order ECM model. Then the accurate estimation of SOC and SOH is achieved based on the High-order Sliding Membrane (HOSM) observer. Also based on the first-order model, [10] establishes the correlation between the mass transfer resistance and the current by analyzing the electrochemical impedance spectrum of the lithium-ion battery. Then an adaptive battery state estimator is proposed to achieve joint estimation for both SOC and the State-of-Available Power (SOAP) of the battery system. [11] achieves accurate SOC estimation by introducing an improved Extended Kalman Filter (EKF) to update the parameters of the first-order ECM model according to the average SOC change. In [12], a recursive least square regression algorithm with forgetting factor is proposed to identify the parameters of a second-order ECM model online. Then the joint estimation of SOH, SOC and the State of Function (SOF) is achieved by EKF. [13][14] implement intelligent searching algorithms such as generic algorithm and Particle Swarm Optimization (PSO) to optimize the parameters of the fractional ECM model. Accordingly, a fractional-order Kalman Filter and a double-fractional-order EKF are proposed respectively to estimate SOC and SOH. The performance of the physics-based methods depends heavily on the accuracy of the physical model. However, due to the complexity and nonlinearity of the battery electrochemical process, existing physical models have inevitable modeling errors. Such modeling errors will consequently lead to state estimation errors, especially under extreme operating conditions such as heavy current charging/discharging and freezing temperature environment.

Data-driven methods try to directly build the mapping from field measurement data, e.g., current, voltage and temperature, to the battery state variables by statistic models [15] or neural networks [16][17]. In [18], real-time measurements of battery voltage, current and ambient temperature are fed into a dynamically-driven recurrent network (DDRN) to estimate the SOC and SOH of the electric vehicle battery. [19][20] also

This work was supported by National Natural Science Foundation of China under Grant 52307121, and also supported by Shanghai Sailing Program under Grant 23YF1419000. (*Corresponding author: Yiyan Li.*)

Zelin Guo, Yiyan Li are with the College of Smart Energy, Shanghai Non-Carbon Energy Conversion and Utilization Institute, and Key Laboratory of Control of Power Transmission and Conversion, Ministry of Education, Shanghai Jiao Tong University, Shanghai, 200240, China. (e-mail: gzl1996@sjtu.edu.cn, yiyan.li@sjtu.edu.cn)

Zheng Yan is with the Key Laboratory of Control of Power Transmission and Conversion, Ministry of Education, and the Shanghai Non-Carbon Energy Conversion and Utilization Institute, Shanghai Jiao Tong University, Shanghai 200240, China. (e-mail: yanz@sjtu.edu.cn).

Mo-Yuen Chow is with the University of Michigan - Shanghai Jiao Tong University Joint Institute, Shanghai Jiao Tong University, Shanghai, 200240, China (email: moyuen.chow@sjtu.edu.cn).

take the voltage and current measurements as the input to a deep feedforward neural network (DFNN) and a stacked long short-term memory network (sLSTM) respectively to estimate SOC, showing better accuracy than unscented Kalman filter on the dynamic stress test, US06 test and other data sets. Data-driven methods are easy to implement and can achieve higher accuracy than physics-based methods due to the strong nonlinear learning ability of the neural nets. However, data-driven methods rely heavily on the quality and quantity of the field measurement data, which impairs the model stability, interpretability and robustness.

In this paper, we try to combine both the advantages of physics-based model and data-driven model by embedding neural networks into the traditional ECM as "*virtual electronic components*". More specifically, three Feedforward Neural Networks (FNNs) are merged into the First-Order ECM to act as the residuals of the Ohm Resistance, Polarization Resistance and Polarization Capacitance respectively. These FNNs will be trained by our proposed iterative offline training strategy based on the field measurement data to dynamically compensate the fitting error of the traditional ECM. Because the neural networks have strong nonlinear fitting ability, the proposed neural-network-embedded ECM (called *hybrid model* in this paper) can better adapt to drastically changing environments and achieve higher accuracy. Consequently, the battery state estimation can be improved based on the more accurate hybrid model.

The main contributions of this paper are twofold:
- We propose a hybrid model structure for lithium-ion batteries by embedding neural networks as "virtual electronic components" into the classical ECM. The neural network modules are trained to fit the residuals of the classical ECM, which can increase the model nonlinear-fitting ability and adaptability and reduce the fitting errors under extreme working conditions.
- We propose an iterative offline training strategy to solve the indirect training problem of the embedded FNN modules. The battery state space model is merged into the traditional Mean Squared Error (MSE) to estimate the terminal voltage prediction accuracy, which serves as a physics-informed loss function to guide the training of the FNNs.

The remainder of this paper is organized as follows. In Section II, the classical ECM used in this paper and the parameter identification method are introduced. In Section III, the proposed hybrid model and the parameter correction method are proposed. In Section IV, a real-world battery dataset is used to verify the accuracy of the hybrid model. Section V concludes this paper.

## II. BASIC ECM AND INITIAL PARAMETER IDENTIFICATION

### A. Basic equivalent circuit model of the lithium-ion battery

Fig.1 shows the First-Order ECM of the lithium-ion battery under discharging condition. The model describes the dynamic polarization characteristics of the battery by a RC circuit, and the basic circuit elements include $R_0, R_D, C_D$. $R_0$ is the ohmic internal resistance, $R_D$ is the polarization resistance, and $C_D$ is the battery capacitance. In this paper, we choose the simplest First-Order ECM as an example to demonstrate how the proposed hybrid model is formulated and trained. Note that such a hybrid modeling strategy can be easily extended to more complicated ECMs such as the Second-Order ECM or the Fractional-Order ECM.

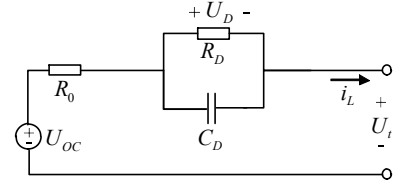

**Fig. 1.** First-Order ECM of the battery

When the battery is discharged, the state space equation of the First-Order ECM can be described as

$$\begin{cases} \dfrac{dU_D}{dt} = \dfrac{i_L}{C_D} - \dfrac{U_D}{R_D C_D} \\ U_t = U_{OC} - U_D - i_L R_0 \end{cases} \quad (1)$$

where $U_D$ is the voltage of the $R_D C_D$ parallel part, $i_L$ is the battery discharging current, $U_t$ is the terminal voltage of the battery, $U_{OC}$ is the open circuit voltage of the battery and is a function of the battery SOC

$$U_{OC} = f(SOC) \quad (2)$$

where $f(\cdot)$ is a polynomial function that can be established by pre-experiments. Because SOC is not measurable, the SOC is calculated by the ampere-hour integration

$$SOC(t) = SOC(0) - \int_0^t \dfrac{i_L dt}{C_b} \quad (3)$$

where $SOC(0)$ is the initial SOC value when the battery starts discharging. $C_b$ is the maximum available capacity of the battery under the current discharging cycle.

### B. Initial parameter identification

In this paper, the parameter identification results of the traditional ECM are called the *initial parameters* (i.e. the parameters that have not been corrected by the FNNs). The initial parameters can be obtained online based on field measurements by using the recursive least squares method with forgetting factor (FFRLS). Details of FFRLS can be found in [21]. The First-Order ECM equation in (1) can be transformed into the least-squares form as

$$y_k = \theta_k \phi_k \quad (4)$$

$$\theta_k = \begin{bmatrix} \dfrac{T - 2R_D C_D}{T + 2R_D C_D} \\ \dfrac{R_0 T + R_D T + 2R_0 R_D C_D}{T + 2R_D C_D} \\ \dfrac{R_0 T + R_D T - 2R_0 R_D C_D}{T + 2R_D C_D} \end{bmatrix}^T \quad (5)$$

$$\phi_k = \begin{bmatrix} -y_{k-1} & I_k & I_{k-1} \end{bmatrix}^T$$

where $y_k$ is the output of the system, i.e. $U_t - U_{OC}$ in this paper. $\theta_k$ is the parameter matrix and $\phi_k$ is the data matrix. $T$ is the sampling interval ($T$=1s in this paper). $I_k$ is the discretized expression of the current $i_L$.

The parameter optimization process of FFRLS is as follows

$$
\begin{cases}
\hat{\theta}_k = \hat{\theta}_{k-1} + K_k(y_k - \phi_k\hat{\theta}_{k-1}) \\
K_k = \dfrac{P_{k-1}\theta_k}{\lambda + \phi_k P_{k-1}\phi_k^T} \\
P_k = \dfrac{(I - K_k\phi_k)P_{k-1}}{\lambda}
\end{cases}
\tag{6}
$$

where $\lambda$ is the forgetting factor, $\hat{\theta}$ is the estimated value of the parameter matrix, $K_k$ is the gain matrix, $P_k$ is the covariance matrix, $I$ is the identity matrix.

FFRLS updates the model parameters at each time step to obtain the initial parameter identification results.

## III. Hybrid Model and Parameter Correction

### A. Neural-Network-Embedded Equivalent Circuit Model

When the battery works in extreme conditions such as freezing temperature or heavy charging/discharging mode, the electrochemical process of the battery becomes rather complex and nonlinear, leading to risks of large fitting errors of the ECMs. In this section, we merge 3 FNN modules (i.e. FNN1, FNN2, FNN3) in to the First-Order ECM to enhance its adaptability and flexibility, as shown in Fig. 2.

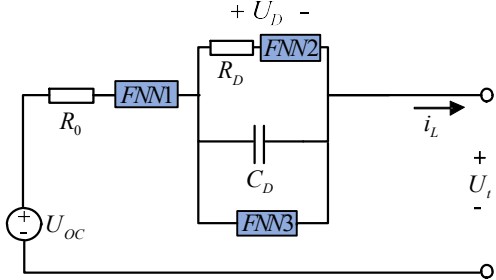

**Fig. 2.** Structure of the hybrid model

The architecture of the 3 FNN modules is shown in Fig. 3. The inputs are field measurements $i_{L,t}$, $U_t$ and $Temp_t$, while the outputs, denoted by $R_{0\_FNN}$, $R_{D\_FNN}$, $C_{D\_FNN}$, are expected to be the corrections for the initial parameters. We expect the FNN modules can correct the initial parameter estimation errors to improve the modeling accuracy. Then the final parameter identification results of the hybrid model can be expressed by

$$
\begin{cases}
\tilde{R}_0 = R_{0\_FFRLS} + R_{0\_FNN} \\
\tilde{R}_D = R_{D\_FFRLS} + R_{D\_FNN} \\
\tilde{C}_D = C_{D\_FFRLS} + C_{D\_FNN}
\end{cases}
\tag{7}
$$

where $\tilde{R}_0$, $\tilde{R}_D$, $\tilde{C}_D$ are the parameter values after correction. Accordingly, the state space equation of the hybrid model can be rewritten as

$$
\begin{cases}
\dfrac{dU_D}{dt} = \dfrac{i_L}{\tilde{C}_D} - \dfrac{U_D}{\tilde{R}_D\tilde{C}_D} \\
U_t = U_{OC} - U_D - i_L\tilde{R}_0
\end{cases}
\tag{8}
$$

Note that in this paper we use the simplest neural network structure, i.e. FNN, to demonstrate the hybrid modeling and training strategy. Researchers can also explore more accurate and efficient network structures to improve the hybrid model performance in the follow-up works.

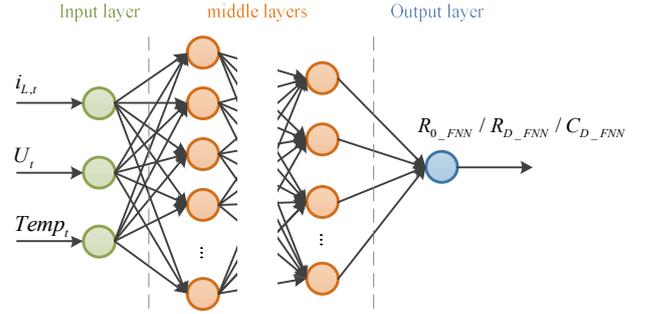

**Fig. 3.** The FNN architecture

### B. Offline training strategy of the hybrid model

The FNN modules cannot be trained directly as a typical regression problem because the outputs, i.e. $R_{0\_FNN}$, $R_{D\_FNN}$, $C_{D\_FNN}$, are not measurable. However, we can calculate the predicted value of the battery terminal voltage $U_t^{pre}$ based on equation (8), and then indirectly estimate the FNN performance by comparing $U_t^{pre}$ with the ground truth $U_t$. following this idea, in this paper we merge the state space equation of the First-Order ECM into the FNN loss function to train the FNN modules.

Specifically, the outputs of the 3 FNN modules are used as the adjusted values of the parameters, and the corrected values of the parameters are calculated according to (7). The corrected values of the parameters are then fed into the battery state space equation to calculate the predicted value of the battery terminal voltage. Finally, the mean square error (MSE) between the predicted values of the hybrid model's terminal voltage and the actual values of the battery's terminal voltage was calculated, and it was used as a new loss function to participate in the backpropagation process of the 3 FNN modules.

To facilitate the program calculation, the state space equation of (8) needs to be discretized as

$$
U_D(k+1) = e^{-\frac{1}{\tilde{R}_D(k)\tilde{C}_D(k)}} \cdot U_D(k) + \tilde{R}_D(k)(1 - e^{-\frac{1}{\tilde{R}_D(k)\tilde{C}_D(k)}}) \cdot i_L(k)
$$
$$
U_t^{pre}(k+1) = U_{OC}(k+1) - i_L(k+1) \cdot \tilde{R}_0(k+1) - U_D(k+1)
\tag{9}
$$

The modified loss function is described as

$$
loss = \frac{1}{n}\sum_{i=1}^{n}(U_{t,i}^{pre} - U_{t,i}^{true})^2
\tag{10}
$$

where $U_{t,i}^{pre}$ is the predicted value of the terminal voltage of the hybrid model at time $i$, $U_{t,i}^{true}$ is the actual value of the terminal voltage of the battery at time $i$, $n$ is the number of sample points.

To summarize, the offline training strategy of the hybrid model is shown in Fig. 4, including the following steps:

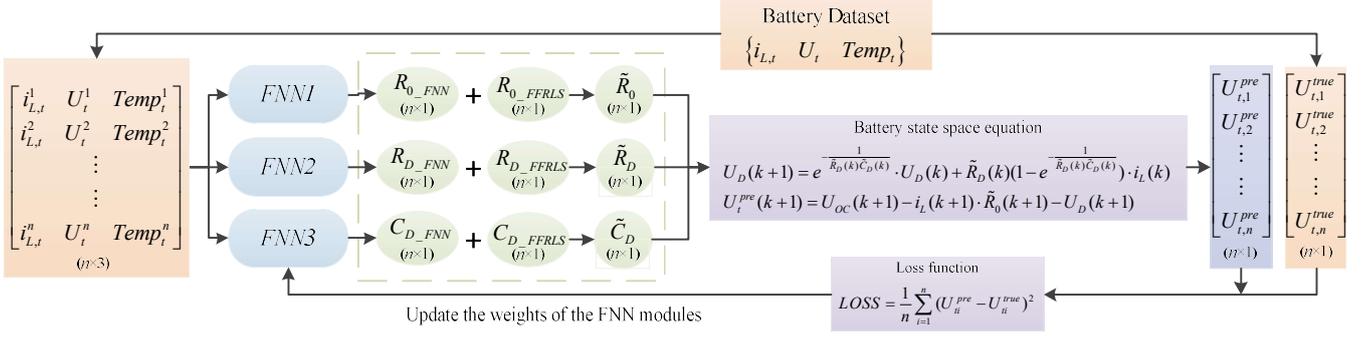

**Fig. 4.** Training strategy of the hybrid model

1) Forward calculation. After the FNNs are initialized, we feed the input variables (i.e. current, voltage and temperature) into the FNNs to obtain their outputs.

2) Parameter correction. The outputs of FNNs are used to correct the initial parameter estimation results, shown as equation (7).

3) Loss calculation. The corrected parameters in step 2) are then fed into the discretized state space equation in (9) to obtain the predicted value of the terminal voltage $U_{t,i}^{pre}$. The MSE loss can be calculated by comparing $U_{t,i}^{pre}$ with the ground truth $U_{t,i}^{true}$.

4) Back propagation. Based on the loss in step 3), the parameters of the FNNs can be updated via the back-propagation process.

5) Stopping criteria. Step 1) - 4) continues iteratively until $U_{t,i}^{pre}$ is close enough to $U_{t,i}^{true}$ with a certain quantitative criteria. Then the FNNs are considered well-trained to fit the ECM residuals, and the training process will end. The parameters obtained from offline training will be applied to online SOC estimation.

Note that the parameters obtained by offline training will not be updated in the online battery SOC estimation stage, due to considerations of the model training cost. Instead, in practice we will retrain the model periodically to update model parameters using the latest field measurement data to guarantee the model accuracy. For example, considering the battery parameters will drift along with the charging/discharging cycles[22] and the self-discharging behaviors[23], we can retrain the FNNs every 50 cycles or every 15 days to update the parameters.

### C. Online SOC estimation for lithium batteries

EKF is a nonlinear extension of the standard Kalman filter[24], and the calculation process is divided into prediction phase and update phase.

In the prediction phase, based on the estimated value $\hat{x}_{k-1}$ of the system state at time $k$-1, the prior estimate $\hat{x}_{k|k-1}$ of the system state at time $k$ and the covariance matrix $P_{k|k-1}$ of the prediction error of the state variable is calculated

$$\hat{x}_{k|k-1} = g(\hat{x}_{k-1}, u_{k-1}) \tag{11}$$

$$P_{k|k-1} = \operatorname{cov}(x_k - \hat{x}_{k|k-1}) = A_{k-1}P_{k-1}A_{k-1}^T + Q_{k-1} \tag{12}$$

where $g(\bullet)$ is the state transition function of the system, $u_{k-1}$ is the input to the system, $x_k$ is the estimate of the state of the system at time $k$, $A_{k-1}$ is the state transition matrix, $P_{k-1}$ is the error covariance matrix at time $k$-1, $Q_{k-1}$ is the covariance matrix of the process noise.

In the correction phase, the Kalman gain matrix $K_k$ is calculated and the state variables and covariance matrix are corrected as

$$K_k = P_{k|k-1}C_k^T(C_k P_{k|k-1}C_k^T + R)^{-1} \tag{13}$$

$$\hat{x}_k = \hat{x}_{k|k-1} + K_k[y_k - h(\hat{x}_{k|k-1}, u_k)] \tag{14}$$

$$P_k = (I - K_k C_k)P_{k|k-1} \tag{15}$$

where $C_k$ is the output matrix, $R$ is the covariance matrix of the observation noise, $y_k$ is the actual output of the system, $h(\bullet)$ is the output equation of the system, $I$ is the identity matrix.

When $P_k$ is minimized, the optimal estimate of the state variable is obtained.

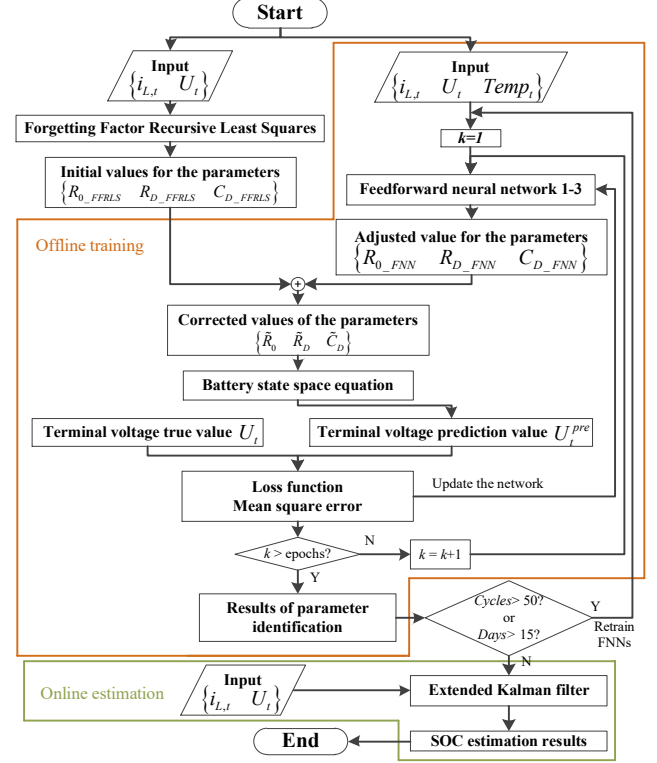

**Fig. 5.** Overall flowchart of the paper

In online application stage, based on the battery equation established by offline training and the real-time measurements of the battery voltage, the online estimation of SOC can be obtained by EKF.

In this paper, MSE and root Mean Square Error (RMSE) [25] are used to calculate the error, which can be described as

$$MSE = \frac{1}{m}\sum_{i=1}^{m}(\tilde{z}_i - z_i)^2 \tag{16}$$

$$RMSE = \sqrt{\frac{1}{m}\sum_{i=1}^{m}(\tilde{z}_i - z_i)^2} \tag{17}$$

where $m$ is the number of sample points, $\tilde{z}_i$ is the predicted value, $z_i$ is the actual value.

The overall framework of this paper is shown in Fig. 5.

## IV. RESULTS AND ANALYSIS

### A. Test case setup

To verify the accuracy of the proposed hybrid model, we set up the test case based on a real-world battery dataset provided by [26] This dataset includes charging/discharging measurements, e.g. current, voltage, cell temperature, from a brand-new Panasonic 18650PF cell under different working conditions. Details of the battery is shown in Table I.

TABLE I
BATTERY INFORMATION

| Parameter | Value |
|---|---|
| Type | 18650PF |
| Rated capacity | 2.9 Ah |
| Minimum capacity | 2.75 Ah |
| Terminal voltage range | 2.5V- 4.2V |
| Temperature of discharge | -20℃ ~ 60℃ |

In this paper, we select 3 different operating modes: HPPC [27], US06 and HWFET [28], at 4 different ambient temperatures: -20℃, -10℃, 0℃, and 10℃ (i.e. $3 \times 4 = 12$ scenarios), to comprehensively evaluate the model performance. As an example, Fig. 6 shows the pulse current curve of the battery when discharging at 10 ℃, and the voltage and SOC profiles when discharging at four temperatures, all under the HPPC operating condition.

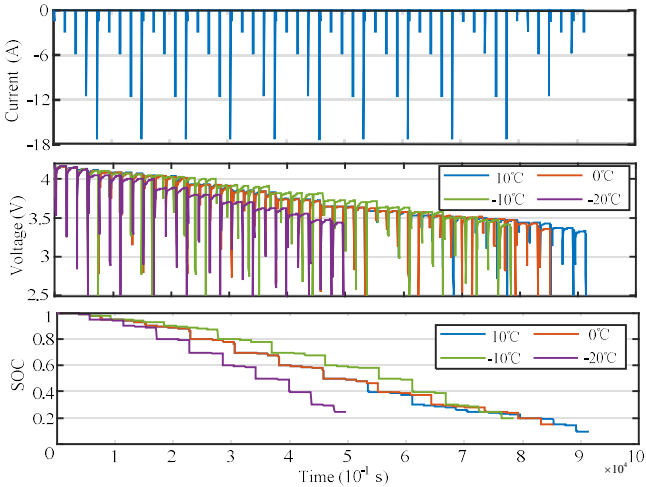

**Fig. 6.** The pulse current curve of the battery when discharging at 10℃, and the changes of the voltage and SOC when discharging at four temperatures

As shown in Fig. 6, the ambient temperature shows significant impact on the discharging performance of the battery: the lower the temperature, the faster the battery SOC curve decreases, resulting in worse discharging performance. Accordingly, the battery parameters are also dependent on the ambient temperature, making the battery parameter identification results at 10 ℃ not applicable to the same battery at -20 ℃.

The configuration of the 3 FNN modules are shown in Table II. Each of the three FNN modules contains an input layer, two intermediate layers and an output layer. Hyperparameters are determined based on trial-and-error strategy to maximize the model performance.

It should be mentioned that in order to speed up the model training, the original measurements are re-sampled at lower frequency (1s in this paper) to obtain the current, voltage and cell temperature data. The SOC values of the battery can be calculated according to (3).

TABLE II
HYPERPARAMETERS FOR THE FNN MODULE

| Hyperparameters | FNN1 | FNN2 | FNN3 |
|---|---|---|---|
| Number of middle-layer nodes | 128/4 | 256/4 | 8/4 |
| Number of training epochs | 200 | 200 | 200 |
| Initial learning rate | 0.001 | 0.01 | 0.01 |
| Optimizer | Adagrad | Adagrad | Adam |

### B. Accuracy verification of the hybrid model

In this section, we first compare the parameter identification results of the classical first-order ECM and the proposed hybrid model. The parameters to be identified include the ohmic internal resistance $R_0$, the polarization resistance $R_D$ and the polarization capacitance $C_D$. The dataset we use is the field measurement data introduced in Section IV.A under HPPC operating condition at 10℃, 0℃, -10℃ and -20℃, respectively.

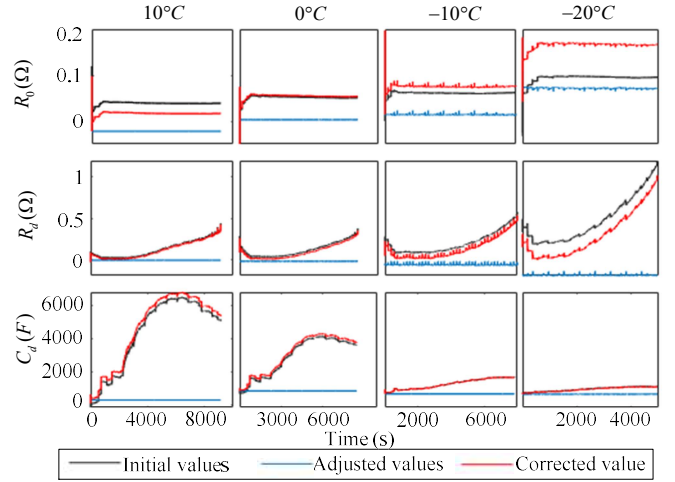

**Fig. 7.** Identification results of battery parameters at different temperatures under HPPC operating mode.

To further validate the model accuracy, we calculate the predicted terminal voltage $U_{t,i}^{pre}$ based on the first-order ECM and the proposed hybrid model with the identified parameters, respectively. In addition to HPPC, we also include another two operating conditions, US06 and HWFET, to comprehensively estimate the model performance. Simulation results are summarized in Fig. 8. We have the following observations:

- Under the HPPC mode, the traditional ECM shows high accuracy at the early stage of the discharging process. However, as the discharging continues, the estimation error of the traditional ECM starts to increase. This is due to the error accumulation effect caused by the recursive FFRLS method and the limited adaptability of the traditional ECM. On the contrary, the proposed hybrid model shows high and stable estimation accuracy during the whole charging process. This is because the FNN modules can dynamically compensate the parameter estimation bias of the traditional ECM based on their strong nonlinear-fitting capability.

- The traditional ECM shows significant fluctuations under US06 and HWFET modes, leading to large estimation errors. Instead, the proposed hybrid model can follow the terminal voltage closely without introducing additional noises. This is because the FNN modules can flexibly learn the fast variation law of voltage and current under US06 and HWFET operating conditions.

- When discharging at -20°C, the traditional ECM yields largest errors under all 3 operating conditions. Because the electrochemical characteristics of the battery become extremely complex and nonlinear, making the traditional

ECM unable to fit. The proposed hybrid model still shows satisfying estimation accuracy due to the strong nonlinear fitting capability brought by the FNN modules.

In Table III, we calculate the Mean Squared Error (MSE) of the traditional ECM and the proposed hybrid model to make a quantitative performance comparison. We can see that the proposed hybrid model shows accuracy improvements under all 3 operating conditions at different ambient temperatures, especially at lower ambient temperatures.

TABLE III
MSE CALCULATION RESULTS

| Operating condition | Temperature | ECM | Hybrid model | Improvement |
|---|---|---|---|---|
| HPPC | 10°C | 0.0053 | 0.0042 | 19.86% |
| | 0°C | 0.0060 | 0.0047 | 21.69% |
| | -10°C | 0.0052 | 0.0029 | 43.85% |
| | -20°C | 0.0096 | 0.0043 | 55.56% |
| US06 | 10°C | 0.0130 | 0.0081 | 37.44% |
| | 0°C | 0.0104 | 0.0018 | 82.53% |
| | -10°C | 0.0145 | 0.0030 | 79.20% |
| | -20°C | 0.0286 | 0.0069 | 76.03% |
| HWFET | 10°C | 0.0019 | 0.0017 | 9.02% |
| | 0°C | 0.0016 | 0.0004 | 77.29% |
| | -10°C | 0.0054 | 0.0027 | 50.36% |
| | -20°C | 0.0104 | 0.0020 | 80.63% |

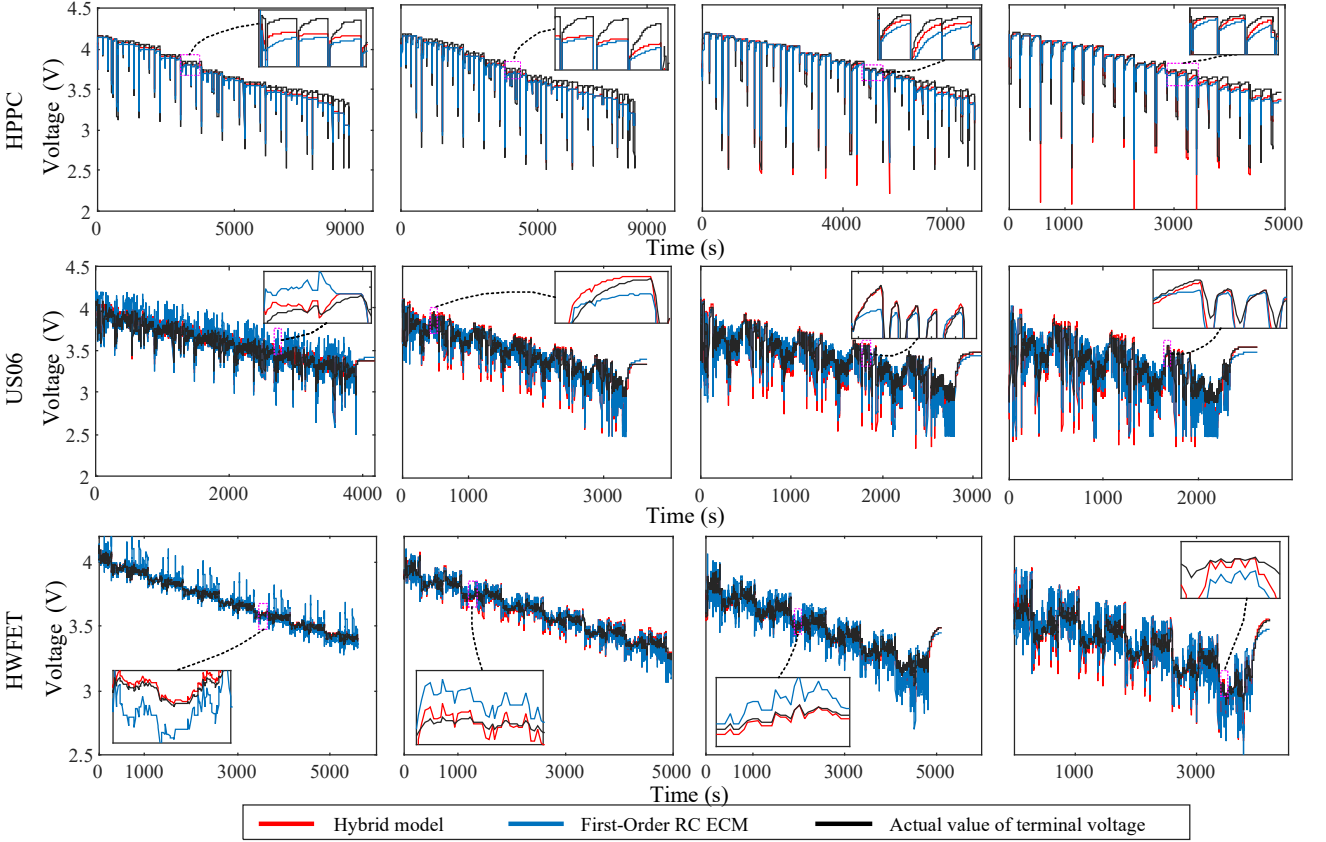

**Fig. 8.** Comparison between the terminal voltage predictions and the ground truth, under different operating conditions and ambient temperatures.

### C. SOC estimation results

Based on the parameter identification results in Section IV.B,

we further implement EKF to estimate the battery SOC during the discharging process to demonstrate the application value of the proposed hybrid model. Table IV shows the average time costs (10℃, 0℃, -10℃, -20℃) of offline training and online SOC estimation under US06 and HWFET operating conditions. We can see that the proposed hybrid model can achieve fast SOC estimation within seconds, which can satisfy the actual engineering application requirements.

TABLE IV
TIME FOR OFFLINE TRAINING AND ONLINE SOC ESTIMATION

|  | Offline training time (s) | Online SOC estimation time (s) |
|---|---|---|
| US06 | 263.940 | 1.325 |
| HWFET | 450.088 | 2.243 |

Fig. 9 shows the SOC estimation results under US06 and HWFET operating conditions, based on the traditional ECM and the proposed hybrid model with their identified parameters, respectively. We have the following observations:

- Overall speaking, the proposed hybrid model shows better SOC estimation accuracy than the traditional ECM at all scenarios, especially at lower temperatures. Such an observation is consistent with the model accuracy comparison results in Section IV.B. Table V provides a quantitative comparison for the SOC estimation accuracy by calculating Root Mean Squared Error (RMSE).

- The traditional ECM suffers from error increase as the discharging process evolves, especially under US06 operating condition. This again shows the advantages of the parameter error compensation and nonlinear fitting capabilities of the proposed hybrid model.

Note that the proposed hybrid model does not always outperform the traditional ECM. For example, the hybrid model shows larger SOC estimation errors at early stages under US06 mode (0℃, -10℃), and at late stages under HWFET mode (10℃, -10℃, -20℃). Because when we train the FNN modules, the loss function is to minimize the global estimation errors during the whole discharging period, during which process some local estimation accuracies are sacrificed.

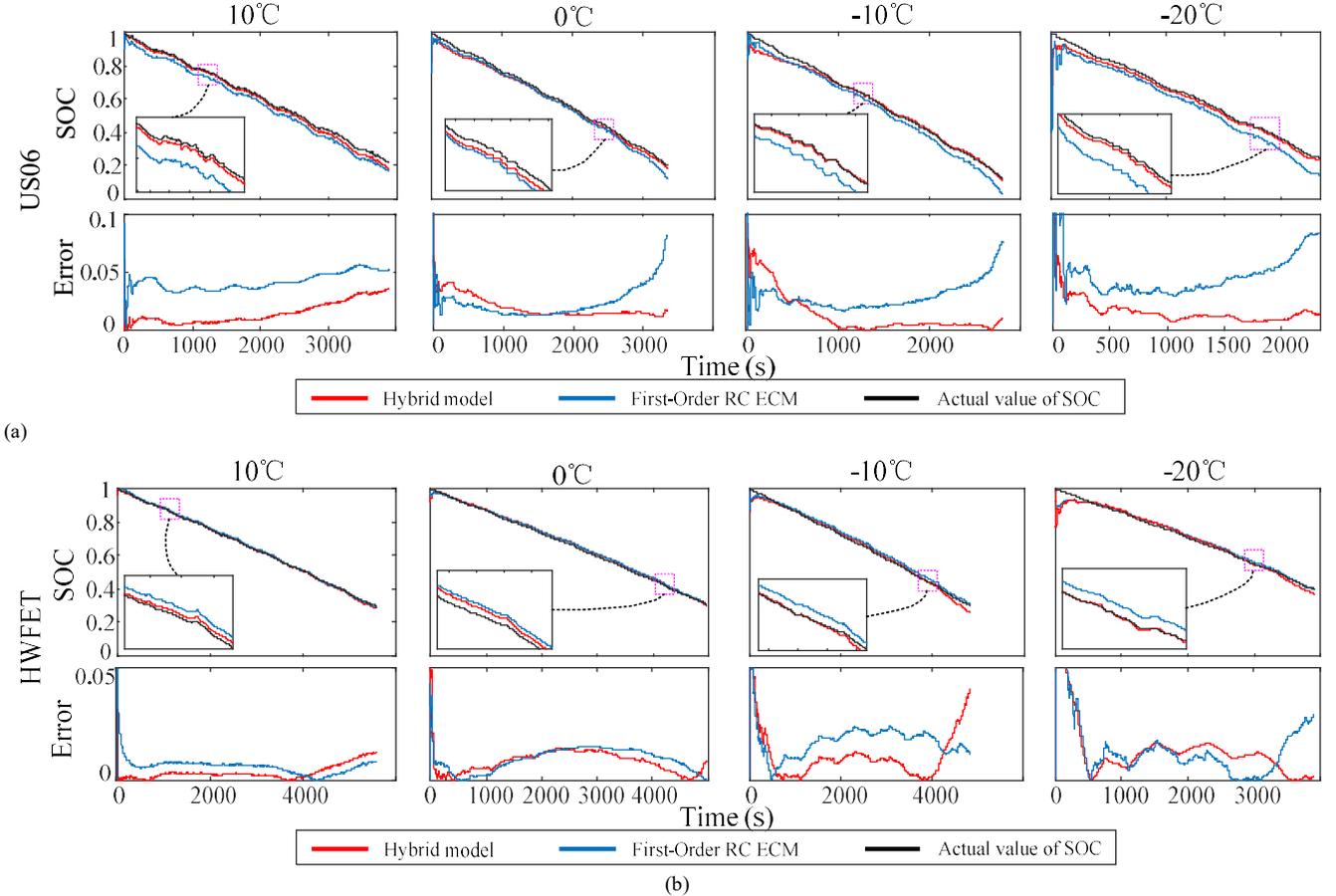

(a)

(b)

**Fig. 9.** SOC estimation results and the estimation errors based on traditional ECM and proposed hybrid model under (a) US06 operating condition, (b) HWFET operating condition.

TABLE V
RMSE COMPARISON OF SOC ESTIMATION RESULTS

| | Temperature | ECM | hybrid model | Improvement |
|---|---|---|---|---|
| US06 | 10℃ | 0.0312 | 0.0183 | 41.31% |
| | 0℃ | 0.0324 | 0.0214 | 33.89% |
| | -10℃ | 0.0374 | 0.0213 | 42.98% |
| | -20℃ | 0.0581 | 0.0209 | 64.03% |
| HWFET | 10℃ | 0.0083 | 0.0078 | 5.30% |
| | 0℃ | 0.0107 | 0.0096 | 9.60% |
| | -10℃ | 0.0215 | 0.0184 | 14.71% |
| | -20℃ | 0.0289 | 0.0206 | 28.67% |

Table V shows the SOC estimation results under two operating conditions, where RMSE is used as the error evaluation index.

## V. CONCLUSION

In this paper, a hybrid model is proposed for lithium-ion batteries to enhance the state estimation accuracy under extreme operating conditions. 3 FNN modules are embedded into the classical first-order ECM, serving as "virtual electronic components" to increase the nonlinear-fitting ability and adaptability. To train the embedded FNN modules, an iterative training process is designed to convert the indirect training problem into a typical regression problem. Particularly, the battery state space equation is merged into the MSE loss to formulate a physics-informed loss function that enables the FNN training. Simulations are conducted based on a real-world battery dataset under 3 different operating conditions at 4 different ambient temperatures. Results show that the proposed method can achieve 9% - 83% error reduction when predicting the battery terminal voltage, and 5% - 64% error reduction when estimating the SOC. Particularly, the error reduction is more significant (44% - 81% for terminal voltage prediction and 15% - 64% for SOC estimation) at lower ambient temperatures (-10℃ and -20℃), indicating that the proposed method can better adapt to extreme operating conditions.

Future work may focus on the fine-tuning of the embedded neural network modules to find more effective structures. One may also try to establish hybrid model based on higher-order ECMs to improve the overall performance.